\documentclass[a4paper,12pt]{article}
\usepackage{amsmath, latexsym, amssymb} \usepackage{euler,eucal}
\usepackage[utf8]{inputenc}
\usepackage[T1]{fontenc}
\usepackage[english]{babel}
\usepackage{tgpagella}

\newcommand{\btau} {\mbox{\boldmath $\tau$}}
\newcommand{\bgam} {\mbox{\boldmath $\gamma$}}

\begin{document}

\title{Squeeze flow of a viscoplastic Bingham medium: an asymptotic solution
\thanks{Partially supported by Federal Programm ``Kadry'' project No. 8226}}

\author{Larisa Muravleva \footnotemark[3]}
\maketitle
\renewcommand{\thefootnote}{\fnsymbol{footnote}}
\footnotetext[3]{Lomonosov Moscow State University, {\tt lvmurav@gmail.com }}
\renewcommand{\thefootnote}{\arabic{footnote}}
\begin{abstract}
We obtain an asymptotic solution for the squeeze flow
of a viscoplastic medium. The standard lubrication-style
expansions of the problem predict plug speed which varies
slowly in the principal flow direction. This variation
implies that the plug region cannot be truly unyielded.
Our solution shows that this region is a pseudo-plug region
in which the leading order equation predict a plug, but
really it is weakly yielded at higher order.
\end{abstract}

\section{Introduction.}

Viscoplastic materials behave as rigid solids, when the
imposed stress is smaller than the yield stress, and
flow as fluids otherwise. The flow field is thus divided
into unyielded (rigid) and yielded (fluid) zones. Two
types of rigid zones are traditionally distinguished: the
stagnation (dead) zones, where the medium is at rest,
and the plug regions (core), where the medium moves
as a rigid body. The surface separating a rigid from a
fluid zone is known as a yield surface. The location and
shape of the latter must be determined as part of the
solution of the flow problem.

In this paper we develop an asymptotic solution for the squeeze flow of a Bingham media.
Squeeze flows are flows in which material is deformed
between two parallel plates approaching each other.
Analysis of viscoplastic fluid flows in geometries with small aspect
ratio has a long history.
Most of the works in the literature deals with the so-called <<lubrication paradox>>
 for yield stress fluids, which refers to the existence or
 not-existence of a true unyielded plug region.
Lipscomb and Denn probably first have shown \cite{LipDenn} that the
usual lubrication approximations  in the squeeze
 flow of Bingham fluids leads to a paradox.
They argued that true rigid plug regions should not exist  in complex geometries, with a
reference to a squeeze flow. The argument is that adoption of classical lubrication
scaling techniques predicts unyielded plug region moves with a speed
which slowly varies in the principal flow direction. This variation implies that
the plug region cannot be truly unyielded. Such regions have been termed pseudo-plug
regions and the boundaries are either pseudo-yield surfaces or fake yield surfaces.
Analytical and numerical methods have shown that
unyielded material may exist in complex geometry: in
axial flows along conduit \cite{WalBit}-\cite{MUR_MTT2009} and
in a flow in a  channel of slowly varying width \cite{FrigRyan}-\cite{MUR_MTT2011}.
Numerous authors made the usual approximations of lubrication theory
for viscoplastic fluids for the axisymmetric geometry and
in the planar geometry \cite{Scott}-\cite{Petr98}.
Detailed review can be found  in \cite{Eng}.   Numerical solution of
an axisymmetric squeeze flow problem can be found in \cite{DonTan}-\cite{MatMit}.
There are three well-known examples of successful application of
asymptotic analysis to the viscoplastic flows in thin-layer problems:
Walton and Bittlestone \cite{WalBit}, Balmforth and Craster \cite{BalmCr},
Frigaard and Ryan \cite{FrigRyan}. In this paper  we try to obtain consistent thin-layer solution
for the squeeze problem of viscoplastic medium.

\section{Problem statement.}

The goal of this study is to obtain a consistent solution for a squeeze
flow of a Bingham viscoplastic fluid for a planar geometry.
We consider the flow of an incompressible Bingham fluid.
The flow is governed by the usual conservation equations of
momentum and mass:
\begin{eqnarray}
\rho \Bigl(\frac{\partial u}{\partial t} +
u \frac{\partial u}{\partial x} + v \frac{\partial u}{\partial y}\Bigr)  &=
- \frac{\partial p}{\partial x} + \frac{\partial \tau_{xx}}{\partial x} +
\frac{\partial \tau_{xy}}{\partial y},\notag\\
\label{mur_1}
\rho \left( \frac{\partial v}{\partial t} +
u \frac{\partial v}{\partial x} + v \frac{\partial v}{\partial y}\right)  &=
- \frac{\partial p}{\partial y} + \frac{\partial \tau_{xy}}{\partial x} +
\frac{\partial \tau_{yy}}{\partial y},\\
\frac{\partial u}{\partial x} + \frac{\partial v}{\partial y} = 0 &\notag
\end{eqnarray}
The $x$-coordinate is aligned with the layer axis, $\rho$ is the density,
$U(u,v)$ is the velocity,
$p$ is the pressure, $\tau_{ij}$ is the deviatoric stress tensor.
The Bingham plastic stresses are related to the  strain rates through
the constitutive equations
\[
\sigma_{ij} = -p\delta_{ij} + \tau_{ij},\quad
\begin{cases}
\tau_{ij} = 2 (\mu  + \frac{\tau_0}{\gamma})e_{ij}, &\mbox{ if }\tau > \tau_0,\\
e_{ij} = 0, &\mbox{ if }\tau \le \tau_0,
\end{cases}
\]
where $\mu$ and $\tau_0$ are respectively
the plastic viscosity and the yield stress, $\gamma_{ij}$ is the strain rate tensor
\[
e_{xx} =  \frac{\partial u}{\partial x},\quad
e_{xy} = \frac12\Bigl(\frac{\partial u}{\partial y} + \frac{\partial v}{\partial x}\Bigr),\quad
e_{xx} =  \frac{\partial v}{\partial y}.
\]

We denote by $\tau$ and $\gamma$ the second invariants of
$\btau$ and $\bgam$, i.e. $\btau = \sqrt{1/2 \tau_{ij} \tau_{ij}}$,
$\bgam = \sqrt{1/2 \gamma_{ij} \gamma_{ij}}$.
The above equations are solved with appropriate boundary conditions:
no slip condition at the plate surfaces
\[
y = H: u = 0,\quad v = -V,
\]
\[
y = -H: u = 0,\quad v = V;
\]
zero tangential and normal stress conditions on the free surface
$x = \pm L$
\[
\sigma_{xx} = -p + \tau_{xx} = 0,\quad
\sigma_{xy} = 0.
\]
We have scaled the lengths in the $x$ and $y$ directions differently,
with the plate half-length $L$ as the horizontal length-scale and
with the half the thickness of the layer $H$ as a characteristic thickness in the $y$-direction,
respectively. We scale the velocity, $u$ and $v$, by $U$ and $U H / L$ ($U H / L = V$)
respectively,
and the time by $L / U$. The pressure is scaled by $\mu U / H$.
The shear-stress components have also been scaled with $\mu U / H$.
The extensional stresses are scaled with $\mu U / L$.
\[
x = L \hat{x},\quad
y = H \hat{y},\quad
u = U \hat{u},\quad
v = \Bigl(\frac{UH}{L}\Bigr) \hat{v},\quad
t = \Bigl(\frac{L}{U}\Bigr) \hat{t},\quad
h = H \hat{h},
\]
\[
\tau_{xx} = \frac{\mu U}{L} \widehat{\tau}_{xx},\quad
\tau_{xy} = \frac{\mu U}{H} \widehat{\tau}_{xy},\quad
\tau_{yy} = \frac{\mu U}{L} \widehat{\tau}_{yy},\quad
p = \frac{\mu U L}{H^2} \widehat{p}.
\]
Substituting the non-dimensional variables into the governing equations \eqref{mur_1},
and dropping the ``tilde'' decoration, we arrive at the following system
of dimensionless equations:
\begin{eqnarray}
\varepsilon Re \Bigl( \frac{\partial u}{\partial t} +
u \frac{\partial u}{\partial x} + v \frac{\partial u}{\partial y}  \Bigr) &=
- \frac{\partial p}{\partial x} + \varepsilon^2 \frac{\partial \tau_{xx}}{\partial x} +
\frac{\partial \tau_{xy}}{\partial y},\notag\\
\label{mur_2}
\varepsilon^2 Re \Bigl( \frac{\partial v}{\partial t} +
u \frac{\partial v}{\partial x} + v \frac{\partial v}{\partial y}  \Bigr) &=
- \frac{\partial p}{\partial y} + \varepsilon^2 \Bigl(\frac{\partial \tau_{xy}}{\partial x} +
\frac{\partial \tau_{yy}}{\partial y}\Bigr),\\
\frac{\partial u}{\partial x} + \frac{\partial v}{\partial y} = 0 &\notag
\end{eqnarray}
In the above equations, the small aspect ratio $\varepsilon$,
the Reynolds number $Re$ and the Bingham number $B$, are defined by
\[
\varepsilon = \frac{H}{L},\quad
Re = \frac{\rho U H}{\mu},\quad
B = \frac{\tau_0 H}{\mu U},
\]

The Bingham number represents the ratio of yield stress
to viscous stress. The equations \eqref{mur_2} are the standard lubrication
equations for any non-Newtonian fluid.
We now assume that $Re \sim O (\varepsilon)$, so that the
inertial terms are of order $O (\varepsilon^3)$,
and will play no part. Equations \eqref{mur_2} are replaced by
\begin{eqnarray}
- \frac{\partial p}{\partial x} + \varepsilon^2 \frac{\partial \tau_{xx}}{\partial x} +
\frac{\partial \tau_{xy}}{\partial y} &= 0,\notag\\
\label{mur_3}
- \frac{\partial p}{\partial y} + \varepsilon^2 \left(\frac{\partial \tau_{xy}}{\partial x} +
\frac{\partial \tau_{yy}}{\partial y}\right) &= 0,\\
\frac{\partial u}{\partial x} + \frac{\partial v}{\partial y} &= 0 \notag
\end{eqnarray}
Exploiting the double symmetry of the plane squeeze flow we consider
the boundary conditions for a part of material contained between the $x$ and $y$
axes and the upper plate.
In the axis of symmetry $y = 0$ (centerline) :
\[
\tau_{xy} = 0,\quad v = 0,
\]
at the contact with the upper plate ($y = 1$):
\[
u = 0,\quad v = -1,
\]
on the axis of symmetry $x = 0$:
\[
u = 0,\quad \tau_{xy} = 0,
\]
on the free  surface $x = 1$:
\[
\sigma_{xx} = - p + \tau_{xx} = 0,\quad
\tau_{xy} = 0.
\]

Hence, provided that $\varepsilon\tau > B$, the stress components are given by
\begin{equation}\label{cons_eq}
\tau_{xx} = 2 \Bigl(1 + \frac{B}{\varepsilon\gamma}\Bigr)\frac{\partial u}{\partial x},\quad
\tau_{xy} = \Bigl(1 + \frac{B}{\varepsilon\gamma}\Bigr)\Bigl(\frac{\partial u}{\partial y} +
\varepsilon^2 \frac{\partial v}{\partial x} \Bigr),\quad
\tau_{yy} = 2 \Bigl(1 + \frac{B}{\varepsilon\gamma}\Bigr)\frac{\partial v}{\partial y},
\end{equation}
\[
\mbox{with }\tau = \frac{1}{\varepsilon} \sqrt{
\tau_{xy}^2 + \frac12 (\tau_{xx}^2 + \tau_{yy}^2) }.
\]
The dimensionless strain rates become
\begin{equation}\label{gam_exp}
\gamma_{xx} = 2 \frac{\partial u}{\partial x},\quad
\gamma_{xy} = \frac{\partial u}{\partial y} + \varepsilon^2 \frac{\partial v}{\partial x},\quad
\gamma_{xx} = 2 \frac{\partial v}{\partial y},
\end{equation}
\[
\mbox{with }
\gamma = \frac{1}{\varepsilon} \Bigr[
\Bigr( \frac{\partial u}{\partial y} + \varepsilon^2 \frac{\partial v}{\partial x}\Bigl)^2
+ 2\varepsilon^2 \Bigr(\frac{\partial u}{\partial x}\Bigl)^2
+ 2\varepsilon^2 \Bigr(\frac{\partial v}{\partial y}\Bigl)^2\Bigl].
\]
We utilize the continuity equations
$\tau_{xx} = - \tau_{yy}$, $\gamma_{xx} = - \gamma_{yy}$ and obtain
\[
\tau = \sqrt{
\frac{1}{\varepsilon^2} \tau_{xy}^2 + \tau_{xx}^2},\quad
\mbox{ with }
\gamma = \sqrt{
\Bigr( \frac{\partial u}{\partial y}\frac{1}{\varepsilon}  + \varepsilon \frac{\partial v}{\partial x}\Bigl)^2
+ 4 \Bigr(\frac{\partial u}{\partial x}\Bigl)^2 }
\]

If the medium belongs to rigidity $\varepsilon\tau < B$, then $\frac{\partial u}{\partial x} =
\frac{\partial u}{\partial y} + \varepsilon^2 \frac{\partial v}{\partial x} =
\frac{\partial v}{\partial y} = 0$.

The boundary conditions become
\begin{equation}\label{BC_0}
u\left.\right|_{y=1} = 0,\quad
v\left.\right|_{y=1} = -1.
\end{equation}

\section{Asymptotic expansions.}

We now solve the equations by introducing an asymptotic expansion.
First we consider shear flow near the plate
for which we may find a solution through
a straightforward expansion of the equations.
To solve the equations we expand the solution into a series:
\[
u = u^0 + \varepsilon u^1 + \varepsilon^2 u^2 \ldots,\quad
v = v^0 + \varepsilon v^1 + \varepsilon^2 v^2 \ldots,\quad
p = p^0 + \varepsilon p^1 + \varepsilon^2 p^2 \ldots,\quad
\tau_{ij} = \tau_{ij}^0 + \varepsilon \tau_{ij}^1 + \varepsilon^2 \tau_{ij}^2 \ldots.
\]
We substitute these expansions into the governing equations \eqref{mur_3}, and separate the
terms into an infinite hierarchy of ascending order in power of  $\varepsilon$.
To leading order, the equations become
\begin{eqnarray}
- \frac{\partial p^0}{\partial x} + \frac{\partial \tau_{xy}^0}{\partial y} = 0,\notag\\
\label{mur_4}
- \frac{\partial p^0}{\partial y} = 0,\\
\frac{\partial u^0}{\partial x} + \frac{\partial v^0}{\partial y} = 0.\notag
 \end{eqnarray}

\subsection{Shear region}

After the solution of the first two equations of the system \eqref{mur_4} we have
\[
p^0 = f_0 (x),\quad \tau_{xy}^0 = y f_0'(x).
\]
Provided the yield stress is exceeded,
the zero-order second invariants of strain rate and stress
are given by
\begin{equation}\label{tens_0}
\tau^0 = |\tau^0_{xy}|,
\gamma^0 = \Bigl| {\frac{\partial u^0}{\partial y}} \Bigr| \frac{1}{\varepsilon}.
\end{equation}
The stress tensor components become
\begin{equation}\label{mur_5}
\tau_{xy}^0 = \frac{\partial u^0}{\partial y} + B \mathrm{sgn} \frac{\partial u^0}{\partial y},\quad
\tau_{xx}^0 = 2 \left( \frac{\partial u^0}{\partial x} +
\frac{B \frac{\partial u^0}{\partial x}}{\Bigl|\frac{\partial u^0}{\partial y}\Bigr| } \right),\quad
\gamma^0 = \frac{1}{\varepsilon} \Bigl| \frac{\partial u^0}{\partial y} \Bigr|.
\end{equation}
From the first relation \eqref{mur_5} and
no-slip boundary conditions \eqref{BC_0} we find
\begin{equation}\label{mur_6}
u^0 = (y^2 - 1) \frac{{f_0}'(x)}{2} + B \mathrm{sgn} \frac{\partial u^0}{\partial y} (1 - y)
\end{equation}

We assume that $\frac{\partial u^0}{\partial y}$  has the following sign:
$\frac{\partial u^0}{\partial y} < 0$ if $xy > 0$
and $\frac{\partial u^0}{\partial y} > 0$
if $xy < 0$.
Let us consider the continuity equation \eqref{mur_4}
that we integrate in $y$ exploiting no-slip boundary conditions \eqref{BC_0}.
We thus find
\[
\frac{\partial u^0}{\partial x} = (y^2 - 1)\frac{f''(x)}{2},\quad
\frac{\partial v}{\partial y} = - (y^2 - 1)\frac{f''(x)}{2},
\]
\[
v^0 = -\frac{f''(x)}{2}\Bigl(\frac{y^3}{3} - y + \frac23 \Bigr) - 1.
\]

We now focus on the first-order approximation. We have
\begin{eqnarray}
-\frac{\partial p^1}{\partial x} + -\frac{\partial \tau_{xy}^1}{\partial y} = 0,\notag\\
-\frac{\partial p^1}{\partial y} = 0,\notag\\
\frac{\partial u^1}{\partial x} + \frac{\partial v^1}{\partial y} = 0,\label{mur_7}\\
\gamma^1 = \frac{1}{\varepsilon}
\Bigr[ \Bigr( \frac{\partial u^0}{\partial y} +
\varepsilon \frac{\partial u^1}{\partial y}\Bigl)^2 \Bigl]^{1/2},\notag\\
\tau_{xy}^1 = \frac{\partial u^1}{\partial y},\quad
 u^1\left.\right|_{y=1} = 0,\quad
 v^1\left.\right|_{y=1} = 0.\notag
\end{eqnarray}
After a simple calculation, we get
\begin{equation*}
p^1 = f_1 (x),\quad
\tau_{xy}^1 = y\cdot f_{1}'(x) + f_2 (x),\quad
u^1 = \frac{f_1 ' (x)}{2}(y^2 - 1) + f_2 (x)(y - 1).
\end{equation*}

\subsection{Plastic region.}

At the leading order, at each $x \in [0,1]$ we have $\tau_{xy}^0 = y \frac{\partial p^0}{\partial x}$.
The $\tau_{xy}^0$ exceeds its maximal value at point $y = 1$ and vanish at point $y = 0$.
Thus, there exists a point $y = y_0$ at which $\tau_{xy}^0 = - B$,
that means $\frac{\partial u^0}{\partial y} = 0$. But according to \eqref{tens_0}
$\tau^0 = B$ and $\gamma^0 = 0$. Thus, at the leading order, the yield condition holds at this point and $y_0$
is the position of pseudo-yield surface. The expression for the velocity can be written as follows:
\[
u_0(x,y) =
\begin{cases}
 \frac{B}{2 y_0}(1 - y_0)^2, & y \in [0,\,y_0],\\
 \frac{B}{2 y_0}\Bigl((1 - y_0)^2 - (y - y_0)^2 \Bigr) , & y \in (y_0,\,1],
\end{cases}
\]
\[
y_0 = \frac{B}{|f'(x)|}.
\]
To determine a pseudo-yield surface $y = y_0 (x)$, 
we use the flow rate for the channel 
\[
\int_{0}^{1} u^0 (x, y)~dy = Q (x).
\]
We find that $y_0 (x)$ is the single root of the Buckingham equation:
\[
y_0^3 - y_0 \Bigl(3 + \frac{6Q}{B} \Bigr) + 2 = 0,\quad Q(x) \equiv x. 
\]
On finding the root, we can determine $y_0$. It is evident that $y_0$ depends on $x$.

It can be seen that this zero-order solution has the characteristic
Bingham-Poiseuille profile of velocity in the sheared layer. But
it is  evident that the plug region is not a true plug region
as the leading order velocity varies in the $x$-direction.
This is the essence of the lubrication paradox for the yield stress fluids.

The source of the problem is revealed in the diagonal components of the stress
\eqref{cons_eq}. From the above expression for $u_0$ we can easily see that
$\frac{\partial u^0}{\partial x}$ will not in general vanish while
$\frac{\partial u^0}{\partial y}$ does. In the domain $|y| \le y_0$
we should consider the higher-order equations and pay respect to
the diagonal stress components.

We assume that the domain occupied by the media can be separated into two subdomains:

1. The external regions are situated near the plates where
the yield criterion is reached and overcome, the shear stress is dominant.
We name it ``shear regions''.

2. The inner region includes centerline, the yield criterion is reached,
the shear stress is smaller and in the centerline equal to zero.
The flow is close to the extensional flow. We call this region ``pseudoplug''
or ``plastic region''.

These regions are separated by an interface represented by
the smooth surface pseudo-plug $y = y_0 (x)$.

Let us consider the domain near centerline of the layer $0 < y < y_0$.
Below the fake yield surface, $y =y_0(x)$, the asymptotic expansion
described above breaks down.
To find an appropriate  solution in  this region we look for a slightly
different set of asymptotic sequences. The principal difference is
in the expansion of the horizontal velocity component:
\begin{equation}\label{u_as}
u = u^0 (x) + \varepsilon u^1 (x,y) + \varepsilon^2 u^2 (x,y) + \ldots
\end{equation}
where the property that $\frac{\partial u}{\partial y} = 0$  at $y_0$
is explicitly built in.
We use the same symbol for the pseudo-plug solution.
When we match the two solutions
we add a subscript $p$ to explicitly distinguish the pseudo-plug solution.

Thus, by introducing an asymptotic expansion, we find
\begin{equation}\label{eq3}
\tau_{xx}^{-1} = \frac{2B}{\gamma^0}\frac{\partial u^0}{\partial x},\quad
\tau_{xy}^0 = \frac{B}{\gamma^0}\frac{\partial u^1}{\partial y},\quad
\tau_{yy}^{-1} = \frac{2B}{\gamma^0}\frac{\partial v^0}{\partial y},
\end{equation}
for $\tau_{-1} > B$, with
\[\gamma^0 = \sqrt{
\Bigr( \frac{\partial u^1}{\partial y}\Bigl)^2
+ 4 \Bigr(\frac{\partial u^0}{\partial x}\Bigl)^2},
\]
and
\begin{equation}\label{eq4}
\tau^{-1} = \sqrt{(\tau_{xy}^0)^2 + (\tau_{xx}^{-1})^2} =
\frac{B}{\gamma^0}\sqrt{
\Bigr( \frac{\partial u^1}{\partial y}\Bigl)^2
+ 4 \Bigr(\frac{\partial u^0}{\partial x}\Bigl)^2} = B.
\end{equation}

The zero-order momentum equations \eqref{mur_5} are still valid
and the boundary conditions are $\frac{\partial u^0}{\partial y} \left.\right|_{y = 0} = 0$,
$v^0 \left.\right|_{y = 0} = 0$, $\tau_{xy}^0 \left.\right|_{y = 0} = 0$.
We have $p^0 (x) = f (x)$ and $\tau_{xy}^0 = y \cdot f' (x)$. From \eqref{eq4} we obtain
\begin{equation}\label{tau-1}
\tau_{xx}^{-1} = \sqrt{ B^2 - (yf'(x))^2}.
\end{equation}

The second relation in \eqref{eq3} implies
 \[
\tau_{xy}^0 \cdot \gamma^0 = B \frac{\partial u^1}{\partial y}
\Rightarrow
y\cdot f'(x) \sqrt{\Bigr( \frac{\partial u^1}{\partial y}\Bigl)^2
+ 4 \Bigr(\frac{\partial u^0}{\partial x}\Bigl)^2} = B \frac{\partial u^1}{\partial y}.
\]
The latter can be solved for $\frac{\partial u^1}{\partial y}$
\[
\frac{\partial u^1}{\partial y} =
\frac{2 y f'(x) \frac{\partial u^0}{\partial x}}{\sqrt{B^2 - y^2(f'(x))^2}} =
-\frac{2y \frac{\partial u^0}{\partial x}}{\sqrt{\frac{B^2}{(f'(x))^2} - y^2}}.
\]
Finally, we have
\[
u^1 = 2 \Bigr(\frac{\partial u^0}{\partial x}\Bigl)
\sqrt{\Bigr(\frac{B^2}{f'(x)}\Bigl)^2 - y^2}  + \varphi_1 (x).
\]

For the first order approximation we have the following equations
\begin{eqnarray}
-\frac{\partial p^1}{\partial x} +
\frac{\partial \tau_{xy}^1}{\partial y} +
\frac{\partial \tau_{xx}^{-1}}{\partial x} &= 0,\notag\\
\label{eq_8}
\frac{\partial }{\partial y} (p^1 + \tau_{xx}^{-1}) = 0,\\
\frac{\partial u^1}{\partial x} +
\frac{\partial v^1}{\partial y} = 0.\notag
\end{eqnarray}
The second equation of system \eqref{eq_8} gives $p^1 = - \tau_{xx}^{-1} + \psi (x)$.
By substituting the last expression in first equation of \eqref{eq_8}
\[
-2 \frac{\partial }{\partial x}\sqrt{B^2 - y^2(f'(x))^2} +
\psi'(x) = \frac{\partial \tau_{xy}^{1}}{\partial y}
\]
After integrating (taking into account that $\tau_{xy}^1\left.\right|_{y=0} = 0$)
\[
\tau_{xy}^{1} = y \psi'(x) + f''
\Bigl[
\frac{B^2}{(f')^2}\arcsin\frac{y}{(B/f')} -
y\sqrt{\frac{B^2}{(f')^2} - y^2}
\Bigr]
\]
To find $v^1$ we use the last equation of the system \eqref{eq_8}
and the boundary conditions
\[
\frac{\partial u^1}{\partial x} =
2 \frac{\partial^2 u^0}{\partial x^2}\sqrt{\frac{B^2}{(f')^2} - y^2} -
\frac{2\frac{\partial u^0}{\partial x} B^2 f''}{(f')^3 \sqrt{\frac{B^2}{(f')^2} - y^2}} + \psi'(x),
\]
\[
v^1 = \arcsin\frac{y}{(B/f')}
\Bigl(
\frac{\partial^2 u^0}{\partial x^2}
\frac{B^2}{(f')^2} -
\frac{2\frac{\partial u^0}{\partial x} B^2 f''}{(f')^3}
\Bigr)
- {\psi_1}'(x) y - \frac{\partial^2 u^0}{\partial x^2}\cdot y \sqrt{\frac{B^2}{(f')^2} - y^2}.
\]

\subsection{Matching}

Write down the value of approximate solutions on the border $y = y_0 (x)$,
index $s$ means shear, $p$ -- plastic.
\[
p^s (x) = f_0 (x) + \varepsilon f_1 (x)+ \dots,\quad
p^p (x) = f (x) + \varepsilon \psi (x)+ \dots \Rightarrow
\]
\[
f_0 (x) = f (x),\quad
f_1 (x) = \psi (x),\quad
f'(x) = - \frac{B}{y_0}.
\]
\[
\tau_{xy}^s (x,y_0) = - B + \varepsilon (y_0 \psi' (x) + g(x) ) + \varepsilon^2\dots,
\]
\[
\tau_{xy}^p (x,y_0) = - B + \varepsilon (y_0 \psi' (x) - \Bigl(\frac{B}{y_0}\Bigr)^{'} y_0^2 \frac{\pi}{2}) + \varepsilon^2\dots = - B + \varepsilon (y_0 \psi' (x) + B y'_0 \frac{\pi}{2}) + \varepsilon^2\dots.
\]
Therefore $g (x) =  B y'_0 \frac{\pi}{2}$.
\[
u^s (x,y_0) =  \frac{B}{2 y_0}(y_0 - 1)^2 +
\varepsilon \Bigl(\frac{y_0^2 - 1}{2} \psi' (x) + B y'_0 \frac{\pi}{2} (y_0 - 1) \Bigr),\quad
u^p (x,y_0) = u_0 (x) + \varepsilon\varphi(x)
\]
From the condition that $u^{s} = u^{p}$ we have the following
\[
u_0 (x) = \frac{B}{2 y_0}(y_0 - 1)^2,\quad
\varphi (x) = \frac{y_0^2 - 1}{2} \psi' (x) + B y'_0 \frac{\pi}{2} (y_0 - 1),\quad
\frac{\partial u_0}{\partial x} = \frac{B y_0' (y_0^2 - 1)}{2 y_0^2}.
\]
The second component of the elocity of zero-order is expressed as follows:
\[
v^s (x, y_0) = -\frac{B y'_0}{2 y_0}\Bigl(- \frac{y_0^3}{3} + y_0 -\frac23 \Bigr) - 1,\quad
v^p (x, y_0) = -y_0 \frac{\partial u_0}{\partial x} = - \frac{B y_0' (y_0^2 - 1)}{2 y_0}.
\]
From the continuity condition of $v^{0p}$ and $v^{0s}$ we obtain
\begin{equation*}
- y_0 \frac{\partial u_0}{\partial x} =
\Bigl(- \frac{y_0^3}{3} + y -\frac23 \Bigr)\frac{f''}{2} - 1,
\end{equation*}
and after minor manipulations we get
\[
B y_0' = \frac{3 y_0^2}{{y_0}^3 - 1}.
\]
Thus for the velocity in plastic and shear regions we
obtain the following expressions:
\[
u^s (x) = - \frac{B}{2 y_0} (y^2 - 1) - B (1 - y) +
\varepsilon\Bigl(
(y^2 - 1)\frac{\psi'(x)}{2} +
(y - 1)\frac{3 \pi y_0^2}{2({y_0}^3 -  1)}
\Bigr),
\]
\[
u^p (x) = \frac{B}{2 y_0} (y_0 - 1)^2 +
\varepsilon \Bigl(
\frac{3(y_0 + 1)} {2({y_0}^2 + y_0 +1)} \sqrt{y_0^2 - y^2} +
\frac{y_0^2 - 1}{2}\psi' + (y_0 - 1)\frac{3 \pi y_0^2}{2({y_0}^3 - 1)}
\Bigr),
\]
the shear stress --
\[
\tau_{xy}^s = - \frac{B}{y_0} y + \varepsilon y \psi' (x) + \varepsilon \frac{3}{y_0^3 - 1} \frac{\pi}{2},
\]
\[
\tau_{xy}^p = - \frac{B}{y_0} y + \varepsilon y \psi' (x) +
\varepsilon \frac{3}{y_0^3 - 1}
(y_0^2 \arcsin \frac{y}{y_0} - y \sqrt{y_0^2 - y^2}).
\]

However, the function $\psi (x)$ and the position of pseudo-yield surface $y_0 (x)$
are still unknown. To find them, we consider the flow rate through the channel:
\[
Q (x) = \int_{0}^{y_0} u^p (x)~dx +
 \int_{y_0}^1 u^s (x)~dx =
\frac{B}{6 y_0} (y_0 - 1)^2(y_0 + 2) + \varepsilon
\Bigl\{
\frac{\psi'}{3}(y_0^3 - 1) +
\frac{B \pi}{2}(y_0^2 - 1)y_0',
\Bigr\}
\]
$Q (x) = Q_0 (x) + \varepsilon Q_1 (x)$. We require that $Q = Q_0$, $Q_1 = 0$.
Thus we obtain the equation for the border $y_0$:
\[
(y_0 - 1)^2(y_0 + 2) = \frac{6 Q y_0}{B},
\]
and the expression for $\psi'$
\[
\frac{\psi'}{3}(y_0^3 - 1) = - \frac{B\pi}{2} y_0'(y_0^2 - 1) \Rightarrow
\psi' = -\frac{9 \pi}{2}\frac{y_0^2 (y_0 + 1)}{(y_0^3 - 1)(y_0^2 + y_0 + 1)}.
\]
Finally, we can determine the pressure
\[
p^s = -B^2 \Bigl(y_0 + \frac{1}{2 y_0^2} \Bigr) -
\varepsilon\frac{3 \pi B}{4}
\Bigl(\ln(y_0^2 + y_0 + 1 ) +
\frac{2}{\sqrt{3}}\arctan\frac{2 y_0 + 1}{\sqrt{3}}
\Bigr) + C,
\]
\[
p^p = -B^2 \Bigl( y_0 + \frac{1}{2 y_0^2} \Bigr) -
\varepsilon\frac{3 \pi B}{4}
\Bigl(
\ln(y_0^2 + y_0 + 1) +
\frac{2}{\sqrt{3}}\arctan\frac{2 y_0 + 1}{\sqrt{3}}
\Bigr) -
\varepsilon B \sqrt{1 - \frac{y}{y_0^2} }  + C.
\]
The constant $C$ can be found from the boundary conditions $p\left.\right|_{x = 1} = p_0$.

\end{document}